\documentclass[aps,prd,twocolumn,superscriptaddress,preprintnumbers,floatfix,nofootinbib,notitlepage,showkeys,showpacs]{revtex4-1}

\usepackage[utf8]{inputenc}

\usepackage{graphicx}
\usepackage{hyperref}
\usepackage{latexsym}
\usepackage{amsmath}
\usepackage{amssymb}
\usepackage{bbm}
\usepackage{ulem}
\usepackage{pdfsync}
\usepackage{epsfig}
\usepackage{epstopdf}
\usepackage{subfigure}
\usepackage{color}
\usepackage{comment}
\usepackage{slashed}
\usepackage{multirow}

\newcommand{\avg}[1]{\langle #1 \rangle}
\begin{document}

\title{Pulsar-timing measurement of the circular polarization of the stochastic gravitational-wave background}

\author{Gabriela Sato-Polito}
\email{gsatopo1@jhu.edu}
\affiliation{Department of Physics and Astronomy, Johns Hopkins University, \\
Baltimore, MD 21218, USA}

\author{Marc Kamionkowski}
\affiliation{Department of Physics and Astronomy, Johns Hopkins University, \\
Baltimore, MD 21218, USA}

\begin{abstract}
Pulsar-timing arrays (PTAs) are in the near future expected to detect a stochastic gravitational-wave background (SGWB) produced by a population of inspiralling super-massive black hole binaries. In this work, we consider a background that can be anisotropic and circularly polarized. We use the expansion of the intensity and the circular polarization in terms of spherical harmonics and the overlap reduction functions for each term in this expansion.We propose an unbiased real-space estimator that can separate the intensity and circular-polarization contributions of the SGWB to pulsar-timing-residual correlations and then validate the estimator on simulated data. We compute the signal-to-noise ratio of a circular-polarization component that has a dipole pattern under different assumptions about the PTA. We find that a nearly-maximal circular-polarization dipole may be detectable, which can aid in determining whether or not the background is dominated by a handful of bright sources.
\end{abstract}

\maketitle

\section{Introduction}
Pulsar-timing arrays (PTAs) are a network of millisecond pulsars that can be used as a gravitational-wave detector on galactic scales \cite{1978SvA....22...36S, 1979ApJ...234.1100D}. The measurement is based on the precise determination of arrival times of radio pulses, which feature spatially correlated fluctuations in the presence of gravitational waves (GWs). The primary source of GWs in the PTA band (roughly $1-100$ nHz) is expected to be a stochastic gravitational-wave background (SGWB) produced by a population of inspiralling super-massive black hole binaries (SMBHBs)\cite{Rajagopal:1994zj,Jaffe:2002rt}. Other sources include cosmic strings \cite{Olmez:2010bi, Sousa:2013aaa, Miyamoto:2012ck, Kuroyanagi:2012jf}, phase transitions \cite{Caprini:2010xv}, and relic GWs from inflation \cite{Starobinsky:1979ty, Zhao:2013bba}. 

Current PTA collaborations include the European PTA (EPTA) \cite{Desvignes:2016yex}, the North American Nanohertz Observatory for Gravitational Waves (NANOGrav) \cite{Brazier:2019mmu}, and the Parkes Pulsar Timing Array (PPTA) \cite{Kerr:2020qdo}, which together form the International Pulsar Timing Array (IPTA) \cite{Perera:2019sca}. While a detection of a SGWB is yet to be claimed, increasingly tight upper limits on the SGWB amplitude are being set and NANOGrav has recently reported evidence for a stochastic red-noise process across all pulsars. These limits have important consequences for the sources of the GW background (see, e.g., Refs. \cite{Taylor:2016ftv} and \cite{Chen:2018znx}).

If the origin of a stochastic background is cosmological or sourced by a large population of distant objects, the standard assumption of statistical isotropy is expected to be a good one. In this regime, the correlation between pulsar timing residuals depends only on their angular separation and is given by the Hellings and Downs curve \cite{1983ApJ...265L..39H}. However, some degree of statistical anisotropy is expected. For instance, the finite number of SMBHBs or the presence of a nearby bright source can lead to an anisotropic background \cite{Ravi:2012bz, Cornish:2013aba, Sesana:2008mz}. Characterizing the anisotropy of the background can therefore be a powerful probe of this astrophysical population. 

Another relevant feature of the SGWB that can be probed with PTAs is its polarization. Analogously to electromagnetism, one can define the Stokes parameters for GWs, and the focus of this work is the circular-polarization component. While the chirality of a background produced by many distant sources is expected to vanish, it may be non-negligible if a reduced number of bright sources dominate, since SMBHBs produce chiral GWs if the orbit is observed face-on. In this scenario, anisotropies in the circular polarization would also be produced, as in the case for intensities. Previous work has addressed the circular polarization of the SWGB in the context of measurements with ground and space-based interferometers \cite{Seto:2006dz, Seto:2006hf, Seto:2007tn, Seto:2008sr}, astrometry \cite{Qin:2018yhy}, and PTAs \cite{Kato:2015bye, Belgacem:2020nda}. 

Ref.~\cite{Kato:2015bye} studied the effect of intensity and circular-polarization anisotropies on the correlation between pulsar timing residuals. They propose a way to infer the spherical-harmonic coefficients of the circular-polarization anisotropy using linear combinations of two pulsar pairs, although {\it a priori} knowledge of the background is required. By describing the pulsar positions in the spherical-harmonic basis, Refs.~\cite{Hotinli:2019tpc} and \cite{Belgacem:2020nda} showed that intensity and circular-polarization anisotropies correspond to even and odd bipolar spherical harmonic multipoles and can therefore be measured independently. However, this bipolar-spherical-harmonic formalism relies on the assumption that one has a complete map of the sky at hand. Since the expansion is on the angular position of the pulsars, this would require a dense and uniform distribution of pulsars, which is incompatible with the current observational capabilities of PTAs.

Here we propose an estimator based on the real-space correlation function, that can naturally separate the intensity and circular-polarization contributions. Since circular polarization is defined as the imaginary part of the cross-correlation between $+$ and $\times$ GW polarizations, it corresponds to an imaginary contribution to the correlation between pulsars in the frequency domain. Hence, we construct an estimator for the real and imaginary parts of the correlation function. We follow a similar approach to Ref.~\cite{Anholm:2008wy}, but generalize the detection statistic to account for intensity and circular polarization anisotropies. Furthermore, we extend the result beyond the weak-signal regime, since any detection of circular-polarization anisotropy will require a level of sensitivity beyond the validity of this approximation.

This paper is organized as follows. We begin by outlining the general effect of GWs on pulsar timing residuals in Sec.~\ref{sec:tim}. In Sec.~\ref{sec:corr} we discuss the imprint of the SGWB on the correlation of timing residuals and define the Stokes parameters for GWs. We present the estimators for intensity and circular polarization in Sec.~\ref{sec:est}, apply this estimator for the simple case of detecting a circular polarization dipole in Sec.~\ref{sec:dipole} and conclude in Sec.~\ref{sec:concl}.

\section{Timing Residuals}\label{sec:tim}
A gravitational wave (GW) propagating between the Earth and a pulsar will change the observed arrival time of pulses. The fractional frequency shift for a given pulsar, which we will label by $a$, at a position $\hat{p}_a$ on the sky, induced by a metric perturbation $h_{ij}(t,\hat{\Omega})$ from a GW propagating in the $\hat{\Omega}$ direction, is given by \cite{1979ApJ...234.1100D, Anholm:2008wy}
\begin{equation}
    z_a(t, \hat{p}_a, \hat{\Omega}) = \frac{1}{2}\frac{\hat{p}^i_a\hat{p}^j_a}{1+\hat{\Omega}\cdot\hat{p}_a}\Delta h_{ij},
\end{equation}
where $\Delta h_{ij} \equiv h_{ij}(t_e,\hat{\Omega}) - h_{ij}(t_p,\hat{\Omega})$ is the difference between the metric perturbation at the solar system barycenter, with coordinates $(t_e, \vec{x}_e)$, and at the pulsar, with coordinates $(t_p, \vec{x}_p)$. We choose a coordinate system in which the origin is at the center of the solar system and the pulsar is some distance $L_a$ away, such that $t_e=t$, $\vec{x}_e = 0$, $t_p=t-L_a$ and $\vec{x}_p = L_a\hat{p}_a$.

In the transverse-traceless gauge, the metric perturbation at each point can be written as the following superposition of plane waves \cite{Allen:1997ad}
\begin{equation}
    h_{ij}(t, \hat{\Omega}) = \sum_{A=+,\times} \int_{-\infty}^{\infty} df \ h_A(f,\hat{\Omega}) e^{A}_{ij}(\hat{\Omega}) e^{2\pi if(t-\hat{\Omega}\cdot \vec{x})},
\end{equation}
where the index $A$ labels the $+$ and $\times$ polarizations and $f$ is the frequency of the GW. The Fourier amplitudes $h_A(f, \hat{\Omega})$ are complex functions that satisfy
\begin{equation}
    h^{*}_A(f, \hat{\Omega}) = h_A(-f, \hat{\Omega}),
\end{equation}
and the polarization tensors $e^{A}_{ij}(\hat{\Omega})$ are given by
\begin{equation}
    \begin{split}
        e^+_{ij}(\hat{\Omega}) &= \hat{m}_i\hat{m}_j - \hat{n}_i\hat{n}_j, \\
        e^{\times}_{ij}(\hat{\Omega}) &= \hat{m}_i\hat{n}_j + \hat{n}_i\hat{m}_j,
    \end{split}
\end{equation}
where $\hat{m}$ and $\hat{n}$ are orthogonal unit vectors perpendicular to $\hat{\Omega}$. 

The total frequency shift induced by a stochastic background comprised of GWs of all frequencies and coming from all directions is then given by
\begin{equation}
\begin{split}
    z_a(t) =& \sum_{A=+,\times} \int_{-\infty}^{\infty} df \int d \hat{\Omega}\ h_A(f,\hat{\Omega}) F^A_a(\hat{\Omega}) e^{-2\pi ift} \\ &\times \left( 1 - e^{2\pi ifL_a(1+\hat{\Omega}\cdot \hat{p}_a)} \right),
    \label{eq:timeres}
    \end{split}
\end{equation}
where $a$ labels each pulsar and the antenna beam pattern is defined as
\begin{equation}
    F^A_a(\hat{\Omega}) = \frac{\hat{p}^i_a \hat{p}^j_a e^A_{ij}(\hat{\Omega})}{2(1+ \hat{\Omega}\cdot \hat{p}_a)}.
\end{equation}

By taking the Fourier transform, we can finally write the timing residual in the frequency domain as
\begin{equation}
    z_a(f) = \int d^2 \hat{\Omega} \left( 1 - e^{2\pi ifL_a(1+\hat{\Omega}\cdot \hat{p}_a)} \right) \sum_A h_A(f, \hat{\Omega}) F^A_a(\hat{\Omega}).
    \label{eq:residual_freq}
\end{equation}
The explicit expressions for the antenna beam patterns are shown in App.~\ref{app:orf}.

\begin{figure*}[t]
    \centering
    \includegraphics[width=0.9\textwidth]{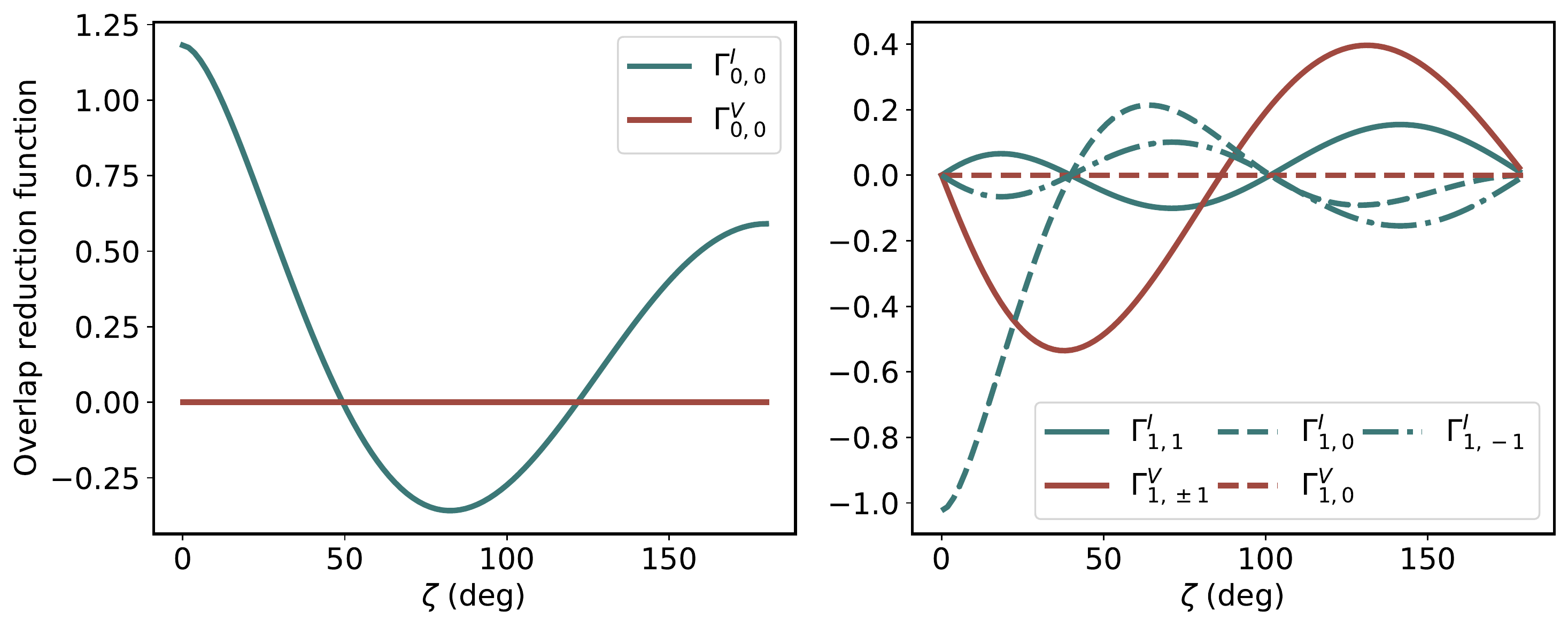}
    \caption{Generalized overlap reduction functions in the computational frame for intensity and circular polarization as a function of pulsar angular separation. The monopole is shown on the left panel and the dipole is shown on the right.}
    \label{fig:ORFs}
\end{figure*}

\section{Correlations}\label{sec:corr}
The presence of a stochastic gravitational wave background induces correlations between the timing residuals of pulsars. Here we consider a background that is Gaussian and stationary, but that can be polarized and anisotropic. In this case, the correlation function between different GW polarizations can be written as
\begin{equation}
    \avg{h^*_A(f, \hat{\Omega}) h_{A'}(f', \hat{\Omega}')} = \delta^2(\hat{\Omega}, \hat{\Omega}') \delta(f-f') \mathcal{P}_{AA'}(f, \hat{\Omega}),
    \label{eq:GWB2PCF}
\end{equation}
where $\mathcal{P}_{AA'}$ is the power spectrum and includes both frequency and angular dependence. 

In analogy with electromagnetism, it is convenient to define the Stokes parameters for GWs as \cite{Kato:2015bye}
\begin{equation}
    \begin{split}
        I &= \frac{1}{2}\avg{|h_+|^2 + |h_{\times}|^2}, \\
        Q &= \frac{1}{2}\avg{|h_+|^2 - |h_{\times}|^2}, \\
        U &= \text{Re}\avg{h^*_+ h_{\times}} = \frac{1}{2}\avg{h^*_+h_{\times} + h^*_{\times}h_+},\\
        V &= \text{Im}\avg{h^*_+ h_{\times}} = \frac{1}{2i}\avg{h^*_+h_{\times} - h^*_{\times}h_+}.
        \label{eq:stokes}
    \end{split}
\end{equation}
With these definitions, the correlation function given in Eq.~\ref{eq:GWB2PCF} can be decomposed as
\begin{equation}
    \mathcal{P}^{AA'} (f, \hat{\Omega}) = \begin{pmatrix} I(f, \hat{\Omega}) + Q(f, \hat{\Omega}) & U(f, \hat{\Omega}) - iV(f, \hat{\Omega}) \\
    U(f, \hat{\Omega}) + iV(f, \hat{\Omega}) & I(f, \hat{\Omega}) - Q(f, \hat{\Omega})
    \end{pmatrix}
\end{equation}
In this work, we will consider only intensity and circular polarization and therefore assume that $Q=U=0$.

With Eq.~\ref{eq:timeres}, we can write the timing residual correlation between pulsars $a$ and $b$. Substituting the definitions of the Stokes parameters in Eq.~\ref{eq:stokes} and keeping only intensity and circular polarization, we get
\begin{widetext}
\begin{equation}
\begin{split}
    \avg{z^*_a(f) z_b(f')} = \int d^2 \hat{\Omega}\ \kappa_{ab}(f, \hat{\Omega}) \delta(f-f') \Big\{I(f, \hat{\Omega})(F^{+*}_{a}F^{+}_{b} + F^{\times*}_{a}F^{\times}_{b})+ iV(f, \hat{\Omega})(F^{+*}_{a}F^{\times}_{b} - F^{\times*}_{a}F^{+}_{b}) \Big\},
\end{split}
\end{equation}
\end{widetext}
where we have defined
\begin{equation}
    \kappa_{ab}(f, \hat{\Omega}) \equiv \left(1 - e^{2\pi ifL_a(1+\hat{\Omega}\cdot \hat{p}_a)}\right)\left( 1 - e^{2\pi ifL_b(1+\hat{\Omega}\cdot \hat{p}_b)} \right).
\end{equation}

We assume that the frequency and angular dependence of the intensity and circular polarization power spectra are separable and expand the angular dependence in spherical harmonics as follows
\begin{equation}
    X(f,\hat{\Omega}) = X(f) \sum_{\ell m} c^X_{\ell m}Y_{\ell m}(\hat{\Omega}),
\end{equation}
where $X = I,V$. Similarly to Ref.~\cite{Kato:2015bye}, we can define the intensity and circular polarization overlap reduction functions (ORFs) as
\begin{equation}
    {}^{(ab)}\Gamma^X = X(f) \sum_{\ell m} c^X_{\ell m} {}^{(ab)}\Gamma^X_{\ell m}
\end{equation}
where we have defined
\begin{equation}
    \begin{split}
        {}^{(ab)}\Gamma^I_{\ell m} &= \int d^2 \hat{\Omega}\ Y_{\ell m}(\hat{\Omega}) \kappa_{ab}(f, \hat{\Omega})  \left(F^{+*}_{a}F^{+}_{b} + F^{\times*}_{a}F^{\times}_{b} \right), \\
        {}^{(ab)}\Gamma^V_{\ell m} &= \int d^2 \hat{\Omega}\ Y_{\ell m}(\hat{\Omega}) \kappa_{ab}(f, \hat{\Omega}) \left(F^{+*}_{a}F^{\times}_{b} - F^{\times*}_{a}F^{+}_{b} \right).
    \end{split}
\end{equation}
Note that we have adopted a slightly different convention for the definitions of the Stokes parameters, but the equations presented here are consistent with Ref.~\cite{Kato:2015bye}. We take the standard assumption that $L_a = L_b$ and that $fL_a \gg 1$ \cite{2018JPhCo...2j5002M, Mingarelli:2014xfa}, such that
\begin{equation}
    \kappa_{ab}(f, \hat{\Omega}) \rightarrow (1+\delta_{ab}),
\end{equation}
and therefore the functions ${}^{(ab)}\Gamma^X_{\ell m}$ are independent of frequency. This approximation is equivalent to neglecting the pulsar term in the cross-correlations, and only including both Earth and pulsar terms in the auto-correlations. Notice that including the pulsar term adds a small change in phase, which corresponds to an imaginary term in the pulsar correlations, and would therefore contaminate the circular polarization measurement.

Substituting the definitions above into the pulsar timing residual correlation, we get
\begin{equation}
     \avg{z^*_a(f) z_b(f')} = \delta(f-f') \left[ I(f) ^{(ab)}\Gamma^I + i V(f) ^{(ab)}\Gamma^V\right].
     \label{eq:res_corr}
\end{equation}
We emphasize here that the quantity $V(f) ^{(ab)}\Gamma^V$ is defined to be real and that circular polarization therefore induces an imaginary contribution to the correlation function.

\begin{figure*}[t]
    \centering
    \includegraphics[width=0.9\textwidth]{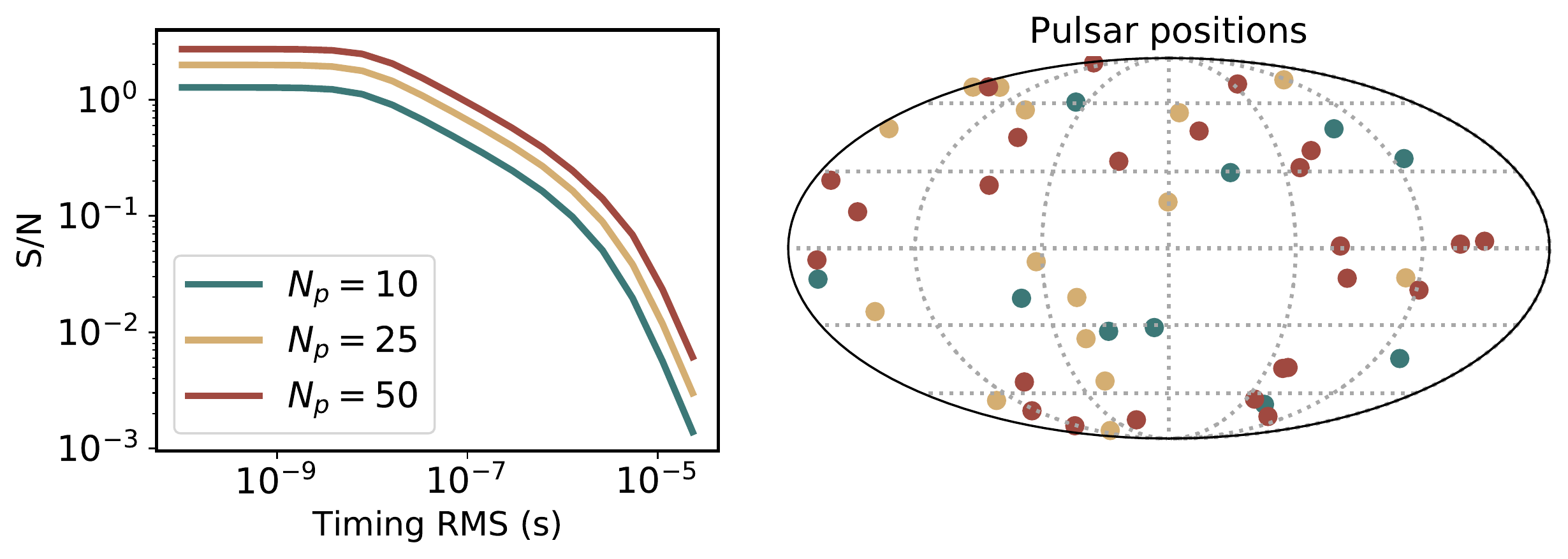}
    \caption{Detectability of the circular polarization dipole as a function of the pulsar white noise. The panel on the left shows the signal-to-noise ratio as a function of the noise RMS for PTAs composed of 10, 25, and 50 pulsars sampled from a uniform spatial distribution. The panel on the right shows the positions of the pulsars on the sky. Note that the inclusion of pulsars is additive: the blue curve on the left panel includes only the blue dots in the map on the right panel, the yellow curve includes the 10 pulsars in blue plus the 15 pulsars in yellow, and the red curve includes all points in the map (blue+yellow+red).}
    \label{fig:SNR_RMS}
\end{figure*}
\section{Estimators for intensity and circular polarization}\label{sec:est}

In this section, we introduce the estimators for intensity and circular polarization of the SGWB. The key result is the signal-to-noise ratio of the circular polarization statistic when the optimal filter and inverse-variance weighting of pulsar pairs are employed \cite{Allen:1997ad, Anholm:2008wy}. We consider the covariance between pulsars in the presence of a non-zero SGWB, thereby accounting for the variance of the background itself \cite{Romano:2020sxq}. 

We assume that the timing residual signal from a given pulsar can be written as a sum of the contributions from the stochastic background and measurement noise as follows
\begin{equation}
    \begin{split}
        s_a(f) = z_a(f) + n_a(f).
    \label{eq:signal}
    \end{split}
\end{equation}
We consider a simplified noise model in which the intrinsic pulsar red noise is omitted, and $n_a$ is assumed to be an uncorrelated white Gaussian noise, characterized by the power spectra
\begin{equation}
   \avg{n^*_a(f)n_b(f')} = \frac{1}{2}\delta_{ab} \delta(f-f')  N_a(|f|),
   \label{eq:noise_power}
\end{equation}
where $N_a(f)=2\sigma^2_a \Delta t$ is the one-sided noise power spectrum of the $a$-th pulsar, $\sigma^2_a$ is its noise variance, and $\Delta t$ is the sampling period. 

The intensity and circular polarization contributions to the background can be identified through the measurement of the real and imaginary parts of the cross-correlation between timing residuals. The estimators for the two components can therefore be written as
\begin{equation}
    \begin{split}
        \hat{I}_{ab} =& \frac{1}{2}\int^{\infty}_{-\infty} df \int^{\infty}_{-\infty} df' \delta_T(f-f') \big[s^*_a(f) s_b(f') \\&+ s_a(f) s^*_b(f')\big] Q^I_{ab}(f'),\\
        \hat{V}_{ab} =& \frac{1}{2i}\int^{\infty}_{-\infty} df \int^{\infty}_{-\infty} df' \delta_T(f-f') \big[s^*_a(f) s_b(f') \\&- s_a(f) s^*_b(f')\big] Q^V_{ab}(f'),
    \label{eq:estimators}
    \end{split}
\end{equation}
where $\delta_T(f) = \sin(\pi f T)/\pi f$, and $Q_I$ and $Q_V$ are filters to be determined. We will focus our discussion on the circular polarization estimator, but note that the optimal estimator for the intensity can be derived in an identical manner.

Our goal is to derive an expression for the filters that maximize the signal-to-noise ratio, and combine the measurement from each pulsar pair in an optimal way. Due to the frequency integration, we require the filter for the intensity estimator to satisfy $Q_I(f) = Q_I(-f)$ and for the circular polarization filter to satisfy $-Q_V(f) = Q_V(-f)$.

The mean of the circular polarization estimator is given by
\begin{equation}
\begin{split}
   \avg{\hat{V}_{ab}} =& T \int_{-\infty}^{\infty} df\ V(f) {}^{(ab)}\Gamma^V Q^V_{ab}(f),
    \label{eq:opt_signal}
\end{split}
\end{equation}
and its covariance is shown in App.~\ref{app:cov}. In order to shorten the notation, we write the variance for a pulsar pair given in Eq.~\ref{eq:cov} as
\begin{equation}
    \sigma^2_{ab} = \frac{T}{2}\int_{-\infty}^{\infty} df \left[Q^V_{ab}(f)\right]^2 \mathcal{C}_{ab}(f),
\end{equation}
where $\mathcal{C}_{ab} \equiv \mathcal{C}_{ab, ab}$.

The optimal filter can be found by defining the inner product in the space of complex valued functions \cite{1997rggr.conf..373A}
\begin{equation}
    (A,B) = \int_{-\infty}^{\infty} df A^*(f) B(f) \mathcal{C}_{ab}(f).
\end{equation}
By writing $\avg{\hat{V}_{ab}}$ and $\sigma^2_{ab}$ as an inner product, it can be shown using the Schwarz inequality that the optimal filter is given by
\begin{equation}
    Q_{ab}(f) = \chi \frac{V(f) {}^{(ab)}\Gamma^V} {\mathcal{C}_{ab}(f)},
\end{equation}
where the constant $\chi$ is chosen so at to set the mean of the circular polarization estimator. The measurement from each pulsar pair, accounting for correlations, can then be optimally combined as follows (see, e.g. Ref~\cite{Romano:2020sxq})
\begin{equation}
    \hat{V}_{\text{opt}} = \frac{\sum\limits_{a}\sum\limits_{b<a}\lambda_{ab} \hat{V}_{ab}}{\sum\limits_{a}\sum\limits_{b<a}\lambda_{ab}},
\end{equation}
where
\begin{equation}
    \lambda_{ab} = \sum\limits_{c}\sum\limits_{d<c} (C^{-1})_{ab,cd}
\end{equation}
and the variance of the optimal estimator is given by
\begin{equation}
    \sigma^2_{\text{opt}} = \frac{1}{\sum\limits_{a}\sum\limits_{b<a}\lambda_{ab}}.
    \label{eq:opt_noise}
\end{equation}

\section{Case Study: Detecting a circular polarization dipole}\label{sec:dipole}
To showcase the proposed circular polarization estimator, we consider a simple scenario in which the gravitational wave background is dominated by a monopole and a dipole that points in the $\hat{z}$ direction. The background is therefore assumed to be
\begin{equation}
        X(f,\hat{\Omega}) = X(f) \left(c^X_{00}Y_{00} + c^X_{10}Y_{10}\right)
\end{equation}
in the cosmic rest-frame. We note that the cross-correlation of pulsars is insensitive to an isotropic circularly polarized background, or, in other words, $^{(ab)}\Gamma^V_{00} = 0$. The inclusion of the $\ell,m=0$ term in the $V$ component is therefore superfluous. We assume the fiducial value of the dipole coefficient to be 15\% of the monopole for both $I$ and $V$.

We further assume that the frequency spectrum for both intensity and circular polarization can be written as
\begin{equation}
    X(f) = \frac{1}{16\pi} A_{\text{X}}^2\left(\frac{f}{f_{1\text{yr}}}\right)^{2\alpha_X -1}.
\end{equation}
We choose the fiducial value of $A_I = A_V = 10^{-15}$ and $\alpha_I=\alpha_V=-2/3$. We focus on estimating the amplitude of the circular polarization of the SGWB. In the dipole toy-model we consider here, the coefficient $c_{10}$ is degenerate with the amplitude of the GW spectrum. Hence, we choose to normalize the  circular polarization estimator such that $\avg{\hat{V}_{ab}} = c^V_{10}A^2_{V}$. The constant $\chi$ is therefore given by
\begin{equation}
    \chi=\frac{c^V_{10}A^2_V}{T}\left(\int_{-\infty}^{\infty} df \frac{ \left(V(f)^{(ab)}\Gamma^V_{10}\right)^2}{\mathcal{C}_{ab}} \right)^{-1}.
\end{equation}

We generate random pulsar positions uniformly distributed on the sky and assume that all pulsars have identical white noise. The integral over frequencies shown throughout this work are, in practice, taken between minimum and maximum frequencies $f_{\text{min}}$ and $f_{\text{max}}$. The frequency range is determined by the total observing time $T$ and the cadence time $\Delta t$ (the time between consecutive pulsar observations). We assume an observing time of 10 yr, which corresponds to $f_{\text{min}} \sim 3\times 10^{-9}$Hz, and a cadence of 2 weeks, resulting in $f_{\text{max}}\sim8\times 10^{-7}$Hz. 

We obtain the signal-to-noise ratio of the circular polarization amplitude in two different ways. First, analytically, by computing the covariance matrix given in Eq.~\ref{eq:cov} and the variance of the optimal estimator using Eq.~\ref{eq:opt_noise}. We then validate our results on simulated timing residuals that we generate using the method described in App.~\ref{app:sims}. The estimator defined in Eq.~\ref{eq:estimators} is then applied to the simulated data, and the signal and noise are given by the mean and standard deviation of the optimal estimator across various realizations of the simulated PTAs. We confirmed that our estimator is unbiased and that the variance matches our predicted value. We therefore choose to present only the predicted signal-to-noise ratio in the results shown in this work.

The signal-to-noise ratio of the circular polarization estimator is shown as a function of noise RMS in Fig.~\ref{fig:SNR_RMS}, for a network of 10, 25, and 50 pulsars. We can see three distinct regimes: weak-signal (noise-dominated), intermediate-signal, and high-signal (SGWB-dominated). The high-signal regime seen here corresponds to a ``cosmic variance'' limit of the circular polarization measurements, which results in a $S/N \sim 3$ for a PTA with 50 pulsars. The second data release by the IPTA~\cite{Perera:2019sca} includes 65 pulsars with values of RMS between $\sim 0.2-14\mu$s, roughly corresponding to the intermediate-noise portion of Fig.~\ref{fig:SNR_RMS} if such a dipolar background were present.

\begin{figure}[t]
    \centering
    \includegraphics[width=0.45\textwidth]{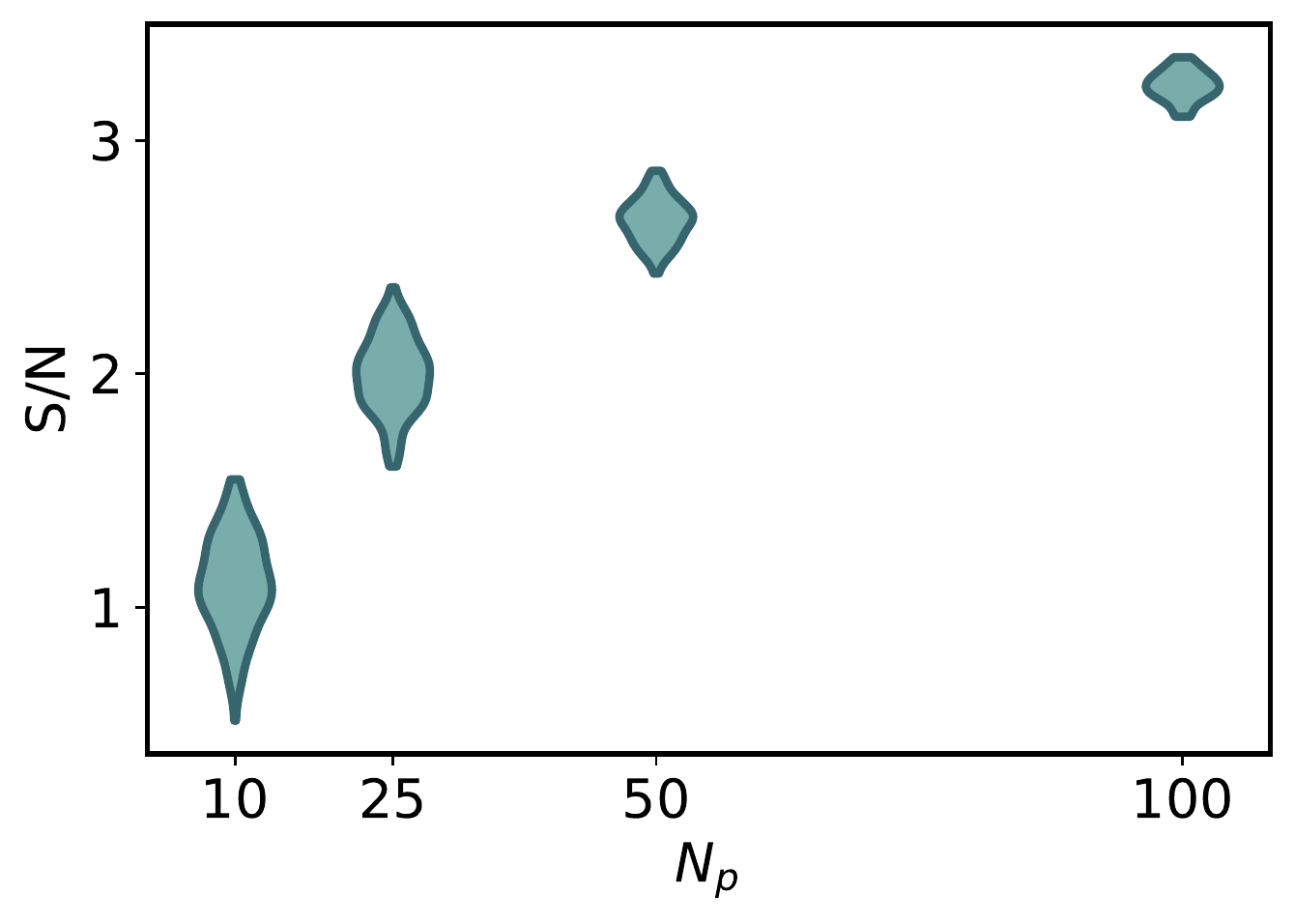}
    \caption{Signal-to-noise ratio of the amplitude of a circular polSarization dipole as a function of the number of pulsars. The distributions correspond to the intrinsic variation in $S/N$ due to different realizations of the pulsar spatial positions. The pulsars were generated assuming a uniform distribution of angles on the sky.}
    \label{fig:SNR_Np}
\end{figure}

The results in Fig.~\ref{fig:SNR_RMS} correspond to a single realization of pulsars on the sky. However, the expected signal-to-noise ratio depends on the positions of the pulsars relative to the SGWB dipole. A different realization of the pulsar positions would therefore lead to a different signal-to-noise estimate. We show in Fig.~\ref{fig:SNR_Np} the distributions of the signal-to-noise ratios for different numbers of pulsars in a PTA. We consider a measurement in the high-signal regime and assume all pulsars have an RMS of 1ns.

As expected, the distribution is the widest for the smallest PTA considered, with a median value around $S/N\sim 1$. As we increase the number of pulsars, the distribution narrows and reaches a signal-to-noise ratio of nearly $S/N\sim 3$ for the largest array considered. In practice, the distribution of pulsars on the sky is not uniform and is clustered around the galactic plane. While the assumption of a uniform spatial distribution of pulsars overestimates the signal-to-noise ratio of the SGWB, Fig.~\ref{fig:SNR_Np} captures a sensible range of values for the dipole toy-model of GW anisotropy.

\section{Conclusions}\label{sec:concl}
Over the last decade, PTA consortia have been gathering increasingly precise timing data and are beginning to dip into an astrophysically interesting region of the SGWB parameter space. The prospect of detection in the near future and the expansion of both telescope-centered programs \cite{Bailes:2018azh, Ng:2017djg} and new PTAs \cite{Susobhanan:2020zmm} motivates us to consider what are the next targets after an isotropic background is measured.

In this work, we address the circular polarization of the SGWB. By isolating the real and imaginary parts of the timing residual correlation, we construct an estimator that can naturally distinguish between the intensity and circular polarization contributions of the SGWB. The estimator is based on extracting the real and imaginary parts of the cross-correlation between pulsar timing residuals. We compute the optimal filter and the combination of pulsar pairs that maximizes the signal-to-noise ratio of the circular polarization component.

As a case study, we consider a simple toy-model in which the SGWB is dominated by a dipole that has 10\% of the power of the monopole and that points in the $\hat{z}$ direction. We then compute the expected signal-to-noise ratio of the circular polarization amplitude under different assumptions about the number of pulsars and their white noise. In order to validate our results, we also apply the proposed estimator on simulated timing residuals and recover the amplitude of the supposed circular polarization dipole.

By varying the pulsar noise RMS, we show where the three (low-signal, intermediate-signal, and high-signal) regimes fall in the dipole toy-model. Under idealized observational assumptions, we show that the dipole can be recovered with a significance $\sim 3\sigma$ in an array that contains 50 pulsars in the SGWB-dominated high-signal regime. Finally, we show how the sensitivity of the PTA to the circular polarization amplitude varies with the locations of the pulsars on the sky. We compute the distribution of signal-to-noise ratios across multiple realizations for increasing numbers of pulsars.

The circular polarization of the SGWB can offer additional information about the astrophysical sources of the background and of more exotic parity-violating physics. In particular, it may be an important tool to distinguish between scenarios in which the background is dominated by multiple distant sources and a smaller number of nearby sources. This contribution is naturally present in the data set of PTAs and therefore extracting the circular polarization component may be an important step in maximizing the information recovered from timing data.

\acknowledgments
We thank Tristan L. Smith, Andrea Lommen, Chiara Mingarelli, and José Luis Bernal for useful discussions. GSP was supported by the National Science Foundation Graduate Research Fellowship under Grant No.\ DGE1746891. This work was supported at Johns Hopkins by NSF Grant No.\ 1818899 and the Simons Foundation.

\appendix
\section{Overlap Functions}\label{app:orf}
In order to compute the ORFs explicitly, we must specify a coordinate system for the pulsar positions and the gravitational wave propagation vector. Following the formalism of Ref.~\cite{Allen:1996gp}, we work in the computational frame, in which one pulsar is defined to be in the $\hat{z}$ direction and the second pulsar in the $x-z$ plane. That is, we work in the following coordinate system:
\begin{equation}
\begin{split}
    \hat{p} &= (0,0,1),\\
    \hat{q} &= (\sin\zeta, 0, \cos\zeta),\\
    \hat{\Omega} &= (\sin\theta\cos\phi, \sin\theta\sin\phi, \cos\theta), \\
    \hat{m} &= (\sin\phi, -\cos\phi, 0), \\
    \hat{n} &= (\cos\theta\cos\phi, \cos\theta\sin\phi, -\sin\theta),
\end{split}
\end{equation}
where $\zeta$ is the angular separation between pulsars $a$ and $b$, and the angles $\theta$ and $\phi$ define the GW propagation direction. In this coordinate system, it can then be shown that the antenna beam functions are given by \cite{Mingarelli:2013dsa}
\begin{equation}
    \begin{split}
        F^{+}_{\hat{p}}(\hat{\Omega}) &= -\frac{1}{2}\left(1-\cos\theta\right), \\
        F^{\times}_{\hat{p}}(\hat{\Omega}) &= 0, \\
        F^{+}_{\hat{q}}(\hat{\Omega}) &= \frac{1}{2}\frac{(\sin\phi\sin\zeta)^2 - \left(\sin\zeta\cos\theta\cos\phi - \sin\theta\cos\zeta\right)^2}{1+\cos\cos\zeta+\sin\theta\sin\zeta\cos\phi}, \\
        F^{\times}_{\hat{q}}(\hat{\Omega}) &= \frac{\left(\sin\phi\sin\zeta\right)\left(\cos\theta\sin\zeta\cos\phi - \sin\theta\cos\zeta\right)}{1+\cos\zeta+\sin\theta\sin\zeta\cos\phi}.
    \end{split}
\end{equation}

From the beam functions above, Ref.\cite{Mingarelli:2013dsa} computed the ORFs for intensity, which we quote below up to the dipole terms

\begin{widetext}
\begin{equation}
    \begin{split}
        {}^{(ab)}\Gamma^I_{00} &= \frac{\sqrt{\pi}}{2}\left[1+\frac{\cos\zeta}{3}+ 4\left(1-\cos\zeta\right)\log\left(\sin\frac{\zeta}{2}\right)\right],\\
        {}^{(ab)}\Gamma^I_{1-1} &= -\frac{1}{2}\sqrt{\frac{\pi}{6}} \left\{ (1+\cos\zeta) + 3(1-\cos\zeta)\left[1+\frac{4}{(1+\cos\zeta)}\log\left(\sin\frac{\zeta}{2}\right)\right] \right\}, \\
        {}^{(ab)}\Gamma^I_{10} &=-\frac{1}{2}\sqrt{\frac{\pi}{3}}\left\{1 +\cos\zeta + 3\left(1-\cos\zeta\right)\left[ 1+\cos\zeta +4\log\left( \sin\frac{\zeta}{2} \right) \right] \right\}, \\
        {}^{(ab)}\Gamma^I_{11} &= -{}^{(ab)}\Gamma^I_{1-1}.
    \end{split}
\end{equation}
\end{widetext}

Similarly, the ORFs for circular polarization were computed in Ref.~\cite{Kato:2015bye}, which we quote up to dipole terms as well:
\begin{equation}
    \begin{split}
        {}^{(ab)}\Gamma^V_{00} &= 0,\\
        {}^{(ab)}\Gamma^V_{1-1} &= -\sqrt{\frac{2\pi}{3}} \sin\zeta\left\{1+3\left(\frac{1-\cos\zeta}{1+\cos\zeta}\right)\log\left(\sin\frac{\zeta}{2}\right)\right\}, \\
        {}^{(ab)}\Gamma^V_{10} &=0,\\
        {}^{(ab)}\Gamma^V_{11} &= {}^{(ab)}\Gamma^V_{1-1}.
    \end{split}
\end{equation}
The ORFs computed in the computational frame must then be rotated back into the cosmic-rest frame.

\section{Covariance}\label{app:cov}
Here we compute the covariance between different pulsar pairs of the intensity and circular polarization estimators. The estimators in Eq.~\ref{eq:estimators} are defined for each pulsar pair, which we label $(ab)$ and $(cd)$. The covariance for the circular polarization estimator is proportional to
\begin{equation}
\begin{split}
    C_{ab, cd} \propto& \avg{\left(s^*_a s_b - s_a s^*_b\right) \left(s^*_c s_d - s_c s^*_d\right)} \\ &-\avg{s^*_a s_b - s_a s^*_b} \avg{s^*_c s_d - s_c s^*_d}.
\end{split}
\end{equation}
For the sake of conciseness, we have denoted $s(f,\hat{p}_a)$ as $s_a$. While the frequency dependence has been omitted, we note that each timing residual signal corresponds to a certain frequency, which must then be integrated over.

Under the assumption that the signal has a Gaussian distribution, we can use Wick's theorem to expand the expression above. Under the assumption that the only non-zero Stokes parameters are I and V, we are left with
\begin{equation}
    \begin{split}
    C_{ab,cd} =& -\frac{1}{4}\int df_a\int df_b\int df_c\int df_d\ Q^V_{ab}(f_b) Q^V_{cd}(f_d) \times \\& \big[\avg{s^*_a s_d}\avg{s_b s^*_c}+\avg{s_a s^*_d}\avg{s^*_b s_c}-\avg{s^*_a s_c}\avg{s_b s^*_d} \\
    &-\avg{s_a s^*_c}\avg{s^*_b s_d}\big].
\end{split}
\end{equation}
We can now substitute Eq.~\ref{eq:signal} for the timing residual and take the correlations, with Eq.~\ref{eq:res_corr} describing the GW signal and Eq.~\ref{eq:noise_power} describing the noise. We find that
\begin{equation}
    \begin{split}
        C_{ab, cd} =& \frac{T}{2}\int df\ Q^V_{ab}(f) Q^V_{cd}(f) \big\{V^2(f)\left( \Gamma^V_{ac}\Gamma^V_{bd} - \Gamma^V_{ad}\Gamma^V_{bc} \right) \\ &+\left(I(f) \Gamma^I_{ac} + \delta_{ac}N_a\right) \left(I(f) \Gamma^I_{bd} + \delta_{bd}N_b\right) \\
        & - \left(I(f) \Gamma^I_{ad} + \delta_{ad}N_a\right)\left(I(f) \Gamma^I_{bc} + \delta_{bc}N_b\right)\big\},
    \label{eq:cov}
    \end{split}
\end{equation}
where the pulsar pairs $(ab)$ and $(cd)$ may be 4 distinct pulsars, or may have one or both pulsars in common. For later convenience, we will call the term inside the brackets $\mathcal{C}_{ab,cd}$, hence
\begin{equation}
        C_{ab,cd} \equiv \frac{T}{2}\int df\ Q^V_{ab}(f) Q^V_{cd}(f) \mathcal{C}_{ab,cd}.
\end{equation}

\section{Simulated PTA}\label{app:sims}
As a proof-of-concept, we apply the estimator proposed above to simulated PTA data \cite{2015PhRvD..91d4048C}. Given a set of $N_p$ pulsars, we begin by defining the correlation matrix ${\bf M}$ using Eq.~\ref{eq:res_corr}, as
\begin{equation}
    \text{M}_{ab} (f) = \frac{T}{2}\left[ I(f) ^{(ab)}\Gamma^I + i V(f) ^{(ab)}\Gamma^V\right],
\end{equation}
where ${\bf M}$ is therefore a $N_p\times N_p$ Hermitian matrix that is a function of frequency. Applying the Cholesky decomposition, we can find the matrix ${\bf H} (f)$ such that
\begin{equation}
    {\bf H}{\bf H}^\text{T} = {\bf M}
\end{equation}
for each frequency. The timing residual of a pulsar $a$ can then be written as
\begin{equation}
    z_a(f) = \sum_{b} H_{ab}(f) w_b(f),
\end{equation}
where $w_a(f) = x_a(f) + iy_a(f)$ is random complex number and both $x_a$ and $y_a$ have zero mean and unit variance. The random variables $w_a(f)$ therefore satisfy
\begin{equation}
    \avg{w^*_a(f)w_b(f')} = \frac{2}{T} \delta_{ab}\delta(f-f').
\end{equation}
We additionally include a source of uncorrelated white Gaussian noise, such that
\begin{equation}
    \avg{n^*_a(f) n_b(f')} = \frac{1}{2} \delta_{ab} \delta(f-f') N_{a},
\end{equation}
where $N_a$ is the noise spectral density. The pulsars simulated in this manner then have the desired cross-correlation
\begin{equation}
    \avg{z^*_a(f)z_b(f')} = \left[ I(f) ^{(ab)}\Gamma^I + i V(f) ^{(ab)}\Gamma^V\right].
\end{equation}

\bibliography{ref.bib}

\providecommand{\href}[2]{#2}\begingroup\raggedright\begin{thebibliography}{10}

\bibitem{1978SvA....22...36S}
M.~V. {Sazhin}, ``{Opportunities for detecting ultralong gravitational
  waves},'' {\em \sovast} {\bfseries 22} (Feb., 1978) 36--38.

\bibitem{1979ApJ...234.1100D}
S.~{Detweiler}, ``{Pulsar timing measurements and the search for gravitational
  waves},'' \href{http://dx.doi.org/10.1086/157593}{{\em \apj} {\bfseries 234}
  (Dec., 1979) 1100--1104}.

\bibitem{Rajagopal:1994zj}
M.~Rajagopal and R.~W. Romani, ``{Ultralow frequency gravitational radiation
  from massive black hole binaries},''
  \href{http://dx.doi.org/10.1086/175813}{{\em Astrophys. J.} {\bfseries 446}
  (1995) 543--549}, \href{http://arxiv.org/abs/astro-ph/9412038}{{\ttfamily
  arXiv:astro-ph/9412038}}.

\bibitem{Jaffe:2002rt}
A.~H. Jaffe and D.~C. Backer, ``{Gravitational waves probe the coalescence rate
  of massive black hole binaries},''
  \href{http://dx.doi.org/10.1086/345443}{{\em Astrophys. J.} {\bfseries 583}
  (2003) 616--631}, \href{http://arxiv.org/abs/astro-ph/0210148}{{\ttfamily
  arXiv:astro-ph/0210148}}.

\bibitem{Olmez:2010bi}
S.~Olmez, V.~Mandic, and X.~Siemens, ``{Gravitational-Wave Stochastic
  Background from Kinks and Cusps on Cosmic Strings},''
  \href{http://dx.doi.org/10.1103/PhysRevD.81.104028}{{\em Phys. Rev. D}
  {\bfseries 81} (2010) 104028},
  \href{http://arxiv.org/abs/1004.0890}{{\ttfamily arXiv:1004.0890
  [astro-ph.CO]}}.

\bibitem{Sousa:2013aaa}
L.~Sousa and P.~P. Avelino, ``{Stochastic Gravitational Wave Background
  generated by Cosmic String Networks: Velocity-Dependent One-Scale model
  versus Scale-Invariant Evolution},''
  \href{http://dx.doi.org/10.1103/PhysRevD.88.023516}{{\em Phys. Rev. D}
  {\bfseries 88} no.~2, (2013) 023516},
  \href{http://arxiv.org/abs/1304.2445}{{\ttfamily arXiv:1304.2445
  [astro-ph.CO]}}.

\bibitem{Miyamoto:2012ck}
K.~Miyamoto and K.~Nakayama, ``{Cosmological and astrophysical constraints on
  superconducting cosmic strings},''
  \href{http://dx.doi.org/10.1088/1475-7516/2013/07/012}{{\em JCAP} {\bfseries
  07} (2013) 012}, \href{http://arxiv.org/abs/1212.6687}{{\ttfamily
  arXiv:1212.6687 [astro-ph.CO]}}.

\bibitem{Kuroyanagi:2012jf}
S.~Kuroyanagi, K.~Miyamoto, T.~Sekiguchi, K.~Takahashi, and J.~Silk,
  ``{Forecast constraints on cosmic strings from future CMB, pulsar timing and
  gravitational wave direct detection experiments},''
  \href{http://dx.doi.org/10.1103/PhysRevD.87.023522}{{\em Phys. Rev. D}
  {\bfseries 87} no.~2, (2013) 023522},
  \href{http://arxiv.org/abs/1210.2829}{{\ttfamily arXiv:1210.2829
  [astro-ph.CO]}}. [Erratum: Phys.Rev.D 87, 069903 (2013)].

\bibitem{Caprini:2010xv}
C.~Caprini, R.~Durrer, and X.~Siemens, ``{Detection of gravitational waves from
  the QCD phase transition with pulsar timing arrays},''
  \href{http://dx.doi.org/10.1103/PhysRevD.82.063511}{{\em Phys. Rev. D}
  {\bfseries 82} (2010) 063511},
  \href{http://arxiv.org/abs/1007.1218}{{\ttfamily arXiv:1007.1218
  [astro-ph.CO]}}.

\bibitem{Starobinsky:1979ty}
A.~A. Starobinsky, ``{Spectrum of relict gravitational radiation and the early
  state of the universe},'' {\em JETP Lett.} {\bfseries 30} (1979) 682--685.

\bibitem{Zhao:2013bba}
W.~Zhao, Y.~Zhang, X.-P. You, and Z.-H. Zhu, ``{Constraints of relic
  gravitational waves by pulsar timing arrays: Forecasts for the FAST and SKA
  projects},'' \href{http://dx.doi.org/10.1103/PhysRevD.87.124012}{{\em Phys.
  Rev. D} {\bfseries 87} no.~12, (2013) 124012},
  \href{http://arxiv.org/abs/1303.6718}{{\ttfamily arXiv:1303.6718
  [astro-ph.CO]}}.

\bibitem{Desvignes:2016yex}
G.~Desvignes {\em et~al.}, ``{High-precision timing of 42 millisecond pulsars
  with the European Pulsar Timing Array},''
  \href{http://dx.doi.org/10.1093/mnras/stw483}{{\em Mon. Not. Roy. Astron.
  Soc.} {\bfseries 458} no.~3, (2016) 3341--3380},
  \href{http://arxiv.org/abs/1602.08511}{{\ttfamily arXiv:1602.08511
  [astro-ph.HE]}}.

\bibitem{Brazier:2019mmu}
A.~Brazier {\em et~al.}, ``{The NANOGrav Program for Gravitational Waves and
  Fundamental Physics},'' \href{http://arxiv.org/abs/1908.05356}{{\ttfamily
  arXiv:1908.05356 [astro-ph.IM]}}.

\bibitem{Kerr:2020qdo}
M.~Kerr {\em et~al.}, ``{The Parkes Pulsar Timing Array project: second data
  release},'' \href{http://dx.doi.org/10.1017/pasa.2020.11}{{\em Publ. Astron.
  Soc. Austral.} {\bfseries 37} (2020) e020},
  \href{http://arxiv.org/abs/2003.09780}{{\ttfamily arXiv:2003.09780
  [astro-ph.IM]}}.

\bibitem{Perera:2019sca}
B.~B.~P. Perera {\em et~al.}, ``{The International Pulsar Timing Array: Second
  data release},'' \href{http://dx.doi.org/10.1093/mnras/stz2857}{{\em Mon.
  Not. Roy. Astron. Soc.} {\bfseries 490} no.~4, (2019) 4666--4687},
  \href{http://arxiv.org/abs/1909.04534}{{\ttfamily arXiv:1909.04534
  [astro-ph.HE]}}.

\bibitem{Taylor:2016ftv}
S.~R. Taylor, J.~Simon, and L.~Sampson, ``{Constraints On The Dynamical
  Environments Of Supermassive Black-hole Binaries Using Pulsar-timing
  Arrays},'' \href{http://dx.doi.org/10.1103/PhysRevLett.118.181102}{{\em Phys.
  Rev. Lett.} {\bfseries 118} no.~18, (2017) 181102},
  \href{http://arxiv.org/abs/1612.02817}{{\ttfamily arXiv:1612.02817
  [astro-ph.GA]}}.

\bibitem{Chen:2018znx}
S.~Chen, A.~Sesana, and C.~J. Conselice, ``{Constraining astrophysical
  observables of Galaxy and Supermassive Black Hole Binary Mergers using Pulsar
  Timing Arrays},'' \href{http://dx.doi.org/10.1093/mnras/stz1722}{{\em Mon.
  Not. Roy. Astron. Soc.} {\bfseries 488} no.~1, (2019) 401--418},
  \href{http://arxiv.org/abs/1810.04184}{{\ttfamily arXiv:1810.04184
  [astro-ph.GA]}}.

\bibitem{1983ApJ...265L..39H}
R.~W. {Hellings} and G.~S. {Downs}, ``{Upper limits on the isotropic
  gravitational radiation background from pulsar timing analysis.},''
  \href{http://dx.doi.org/10.1086/183954}{{\em \apjl} {\bfseries 265} (Feb.,
  1983) L39--L42}.

\bibitem{Ravi:2012bz}
V.~Ravi, J.~S.~B. Wyithe, G.~Hobbs, R.~M. Shannon, R.~N. Manchester, D.~R.~B.
  Yardley, and M.~J. Keith, ``{Does a 'stochastic' background of gravitational
  waves exist in the pulsar timing band?},''
  \href{http://dx.doi.org/10.1088/0004-637X/761/2/84}{{\em Astrophys. J.}
  {\bfseries 761} (2012) 84}, \href{http://arxiv.org/abs/1210.3854}{{\ttfamily
  arXiv:1210.3854 [astro-ph.CO]}}.

\bibitem{Cornish:2013aba}
N.~J. Cornish and A.~Sesana, ``{Pulsar Timing Array Analysis for Black Hole
  Backgrounds},'' \href{http://dx.doi.org/10.1088/0264-9381/30/22/224005}{{\em
  Class. Quant. Grav.} {\bfseries 30} (2013) 224005},
  \href{http://arxiv.org/abs/1305.0326}{{\ttfamily arXiv:1305.0326 [gr-qc]}}.

\bibitem{Sesana:2008mz}
A.~Sesana, A.~Vecchio, and C.~N. Colacino, ``{The stochastic gravitational-wave
  background from massive black hole binary systems: implications for
  observations with Pulsar Timing Arrays},''
  \href{http://dx.doi.org/10.1111/j.1365-2966.2008.13682.x}{{\em Mon. Not. Roy.
  Astron. Soc.} {\bfseries 390} (2008) 192},
  \href{http://arxiv.org/abs/0804.4476}{{\ttfamily arXiv:0804.4476
  [astro-ph]}}.

\bibitem{Seto:2006dz}
N.~Seto, ``{Quest for circular polarization of gravitational wave background
  and orbits of laser interferometers in space},''
  \href{http://dx.doi.org/10.1103/PhysRevD.75.061302}{{\em Phys. Rev. D}
  {\bfseries 75} (2007) 061302},
  \href{http://arxiv.org/abs/astro-ph/0609633}{{\ttfamily
  arXiv:astro-ph/0609633}}.

\bibitem{Seto:2006hf}
N.~Seto, ``{Prospects for direct detection of circular polarization of
  gravitational-wave background},''
  \href{http://dx.doi.org/10.1103/PhysRevLett.97.151101}{{\em Phys. Rev. Lett.}
  {\bfseries 97} (2006) 151101},
  \href{http://arxiv.org/abs/astro-ph/0609504}{{\ttfamily
  arXiv:astro-ph/0609504}}.

\bibitem{Seto:2007tn}
N.~Seto and A.~Taruya, ``{Measuring a Parity Violation Signature in the Early
  Universe via Ground-based Laser Interferometers},''
  \href{http://dx.doi.org/10.1103/PhysRevLett.99.121101}{{\em Phys. Rev. Lett.}
  {\bfseries 99} (2007) 121101},
  \href{http://arxiv.org/abs/0707.0535}{{\ttfamily arXiv:0707.0535
  [astro-ph]}}.

\bibitem{Seto:2008sr}
N.~Seto and A.~Taruya, ``{Polarization analysis of gravitational-wave
  backgrounds from the correlation signals of ground-based interferometers:
  Measuring a circular-polarization mode},''
  \href{http://dx.doi.org/10.1103/PhysRevD.77.103001}{{\em Phys. Rev. D}
  {\bfseries 77} (2008) 103001},
  \href{http://arxiv.org/abs/0801.4185}{{\ttfamily arXiv:0801.4185
  [astro-ph]}}.

\bibitem{Qin:2018yhy}
W.~Qin, K.~K. Boddy, M.~Kamionkowski, and L.~Dai, ``{Pulsar-timing arrays,
  astrometry, and gravitational waves},''
  \href{http://dx.doi.org/10.1103/PhysRevD.99.063002}{{\em Phys. Rev. D}
  {\bfseries 99} no.~6, (2019) 063002},
  \href{http://arxiv.org/abs/1810.02369}{{\ttfamily arXiv:1810.02369
  [astro-ph.CO]}}.

\bibitem{Kato:2015bye}
R.~Kato and J.~Soda, ``{Probing circular polarization in stochastic
  gravitational wave background with pulsar timing arrays},''
  \href{http://dx.doi.org/10.1103/PhysRevD.93.062003}{{\em Phys. Rev. D}
  {\bfseries 93} no.~6, (2016) 062003},
  \href{http://arxiv.org/abs/1512.09139}{{\ttfamily arXiv:1512.09139 [gr-qc]}}.

\bibitem{Belgacem:2020nda}
E.~Belgacem and M.~Kamionkowski, ``{Chirality of the gravitational-wave
  background and pulsar-timing arrays},''
  \href{http://dx.doi.org/10.1103/PhysRevD.102.023004}{{\em Phys. Rev. D}
  {\bfseries 102} no.~2, (2020) 023004},
  \href{http://arxiv.org/abs/2004.05480}{{\ttfamily arXiv:2004.05480
  [astro-ph.CO]}}.

\bibitem{Hotinli:2019tpc}
S.~C. Hotinli, M.~Kamionkowski, and A.~H. Jaffe, ``{The search for anisotropy
  in the gravitational-wave background with pulsar-timing arrays},''
  \href{http://dx.doi.org/10.21105/astro.1904.05348}{{\em Open J. Astrophys.}
  {\bfseries 2} no.~1, (2019) 8},
  \href{http://arxiv.org/abs/1904.05348}{{\ttfamily arXiv:1904.05348
  [astro-ph.CO]}}.

\bibitem{Anholm:2008wy}
M.~Anholm, S.~Ballmer, J.~D.~E. Creighton, L.~R. Price, and X.~Siemens,
  ``{Optimal strategies for gravitational wave stochastic background searches
  in pulsar timing data},''
  \href{http://dx.doi.org/10.1103/PhysRevD.79.084030}{{\em Phys. Rev. D}
  {\bfseries 79} (2009) 084030},
  \href{http://arxiv.org/abs/0809.0701}{{\ttfamily arXiv:0809.0701 [gr-qc]}}.

\bibitem{Allen:1997ad}
B.~Allen and J.~D. Romano, ``{Detecting a stochastic background of
  gravitational radiation: Signal processing strategies and sensitivities},''
  \href{http://dx.doi.org/10.1103/PhysRevD.59.102001}{{\em Phys. Rev. D}
  {\bfseries 59} (1999) 102001},
  \href{http://arxiv.org/abs/gr-qc/9710117}{{\ttfamily arXiv:gr-qc/9710117}}.

\bibitem{2018JPhCo...2j5002M}
C.~M.~F. {Mingarelli} and A.~B. {Mingarelli}, ``{Proving the short-wavelength
  approximation in Pulsar Timing Array gravitational-wave background
  searches},'' \href{http://dx.doi.org/10.1088/2399-6528/aae06d}{{\em Journal
  of Physics Communications} {\bfseries 2} no.~10, (Oct., 2018) 105002},
  \href{http://arxiv.org/abs/1806.06979}{{\ttfamily arXiv:1806.06979
  [astro-ph.IM]}}.

\bibitem{Mingarelli:2014xfa}
C.~M.~F. Mingarelli and T.~Sidery, ``{Effect of small interpulsar distances in
  stochastic gravitational wave background searches with pulsar timing
  arrays},'' \href{http://dx.doi.org/10.1103/PhysRevD.90.062011}{{\em Phys.
  Rev. D} {\bfseries 90} no.~6, (2014) 062011},
  \href{http://arxiv.org/abs/1408.6840}{{\ttfamily arXiv:1408.6840
  [astro-ph.HE]}}.

\bibitem{Romano:2020sxq}
J.~D. Romano, J.~S. Hazboun, X.~Siemens, and A.~M. Archibald,
  ``{Common-spectrum process versus cross-correlation for gravitational-wave
  searches using pulsar timing arrays},''
  \href{http://dx.doi.org/10.1103/PhysRevD.103.063027}{{\em Phys. Rev. D}
  {\bfseries 103} no.~6, (2021) 063027},
  \href{http://arxiv.org/abs/2012.03804}{{\ttfamily arXiv:2012.03804 [gr-qc]}}.

\bibitem{1997rggr.conf..373A}
B.~{Allen}, ``{The Stochastic Gravity-Wave Background: Sources and
  Detection},'' in {\em Relativistic Gravitation and Gravitational Radiation},
  J.-A. {Marck} and J.-P. {Lasota}, eds., p.~373.
\newblock Jan., 1997.
\newblock \href{http://arxiv.org/abs/gr-qc/9604033}{{\ttfamily
  arXiv:gr-qc/9604033 [gr-qc]}}.

\bibitem{Bailes:2018azh}
M.~Bailes {\em et~al.}, ``{MeerTime - the MeerKAT Key Science Program on Pulsar
  Timing},'' \href{http://dx.doi.org/10.22323/1.277.0011}{{\em PoS} {\bfseries
  MeerKAT2016} (2018) 011}, \href{http://arxiv.org/abs/1803.07424}{{\ttfamily
  arXiv:1803.07424 [astro-ph.IM]}}.

\bibitem{Ng:2017djg}
{\bfseries CHIME Pulsar} Collaboration, C.~Ng, ``{Pulsar science with the CHIME
  telescope},'' \href{http://dx.doi.org/10.1017/S1743921317010638}{{\em IAU
  Symp.} {\bfseries 337} (2017) 179--182},
  \href{http://arxiv.org/abs/1711.02104}{{\ttfamily arXiv:1711.02104
  [astro-ph.IM]}}.

\bibitem{Susobhanan:2020zmm}
A.~Susobhanan {\em et~al.}, ``{pinta: The uGMRT Data Processing Pipeline for
  the Indian Pulsar Timing Array},''
  \href{http://dx.doi.org/10.1017/pasa.2021.12}{{\em Publ. Astron. Soc.
  Austral.} {\bfseries 38} (2021) E017},
  \href{http://arxiv.org/abs/2007.02930}{{\ttfamily arXiv:2007.02930
  [astro-ph.IM]}}.

\bibitem{Allen:1996gp}
B.~Allen and A.~C. Ottewill, ``{Detection of anisotropies in the gravitational
  wave stochastic background},''
  \href{http://dx.doi.org/10.1103/PhysRevD.56.545}{{\em Phys. Rev. D}
  {\bfseries 56} (1997) 545--563},
  \href{http://arxiv.org/abs/gr-qc/9607068}{{\ttfamily arXiv:gr-qc/9607068}}.

\bibitem{Mingarelli:2013dsa}
C.~M.~F. Mingarelli, T.~Sidery, I.~Mandel, and A.~Vecchio, ``{Characterizing
  gravitational wave stochastic background anisotropy with pulsar timing
  arrays},'' \href{http://dx.doi.org/10.1103/PhysRevD.88.062005}{{\em Phys.
  Rev. D} {\bfseries 88} no.~6, (2013) 062005},
  \href{http://arxiv.org/abs/1306.5394}{{\ttfamily arXiv:1306.5394
  [astro-ph.HE]}}.

\bibitem{2015PhRvD..91d4048C}
S.~J. {Chamberlin}, J.~D.~E. {Creighton}, X.~{Siemens}, P.~{Demorest},
  J.~{Ellis}, L.~R. {Price}, and J.~D. {Romano}, ``{Time-domain implementation
  of the optimal cross-correlation statistic for stochastic gravitational-wave
  background searches in pulsar timing data},''
  \href{http://dx.doi.org/10.1103/PhysRevD.91.044048}{{\em \prd} {\bfseries 91}
  no.~4, (Feb., 2015) 044048}, \href{http://arxiv.org/abs/1410.8256}{{\ttfamily
  arXiv:1410.8256 [astro-ph.IM]}}.

\end{thebibliography}\endgroup
\bibliographystyle{utcaps}
\bigskip

\end{document}